\documentclass[12pt]{iopart}

\usepackage{graphicx} %%%%% needed to get psfrag txt to scale correctly?
\usepackage[scanall]{psfrag}
\usepackage{latexsym}
\usepackage{amsopn}
\usepackage{iopams} % AMS fonts etc. for IOP jnls

\makeatletter

\let\IfUndefined\@ifundefined

\newtheorem{theorem}{Theorem}[section]
\newtheorem{lemma}[theorem]{Lemma}
\newtheorem{corollary}[theorem]{Corollary}
\newtheorem{proposition}[theorem]{Proposition}
\newtheorem{definition}{Definition}[section]

\newcommand\atendofproof{{\ifvmode\indent\fi
	\unskip\nobreak\hfil\penalty50\vadjust{\penalty500}%
	\hskip2em\hbox{}\nobreak\hfil\hbox{\vrule height1.5ex width .55em}%
	\parfillskip=0pt\finalhyphendemerits=0\par}}
\IfUndefined{qed}{\let\qed=\atendofproof % an AMS-TeX'ism
	}{\let\atendofproof\qed % stick with \qed if already defined
	}

\long\def\titledpar#1{\par\titledparindent{\titledparfont #1}}
\let\titledparindent\indent
\let\titledparfont\bf
\newenvironment{remark}{\begin{quotation}\titledpar{Remark.}}{
						\end{quotation}\par}
		{\end{enumerate}
		 \end{quotation}}

\newenvironment{proof}{\ifvmode \else\par\fi\nobreak\smallskip\nobreak
		\@ifnextchar[\@proofarg{\titledpar{Proof. }}}{\atendofproof\medbreak}
\def\@proofarg[#1]{\titledpar{Proof #1. }}

\IfUndefined{selectfont}{%
	\def\mydefloglike#1#2#3{%
		\expandafter\def\csname#1\endcsname{\mathop{\rm#2}#3}}
}{%
	\def\mydefloglike#1#2#3{\expandafter\def\csname#1\endcsname
		{\mathop{\mathrm{#2}}#3}}
}

	\let\le\leq
	\let\ge\geq
	\@ifundefined{lvert}{%
		\newcommand{\enorm}[1]{{\,\left\Vert#1\right\Vert}}
		
	}{%
		\newcommand{\enorm}[1]{{\,\left\lVert#1\right\rVert}}
		
	}
	
	\IfUndefined{implies}{\newcommand{\implies}{\Rightarrow}}{}
	
	\newbox\@defeqbox
	\setbox\@defeqbox=\hbox{\footnotesize def}
	\newcommand\defeq{\mathop{=}\limits^{\copy\@defeqbox}}
	
	\newcommand\spdot{\,\cdot\,}
	\newcommand{\seq}[2]{\ifmmode \{{#1}_{#2}\}\else $\{#1_{#2}\}$\fi}
	\let\sub=\subset
	\def\@psup{\mathop{\mathop\supset\limits_{\neq}}}
	\IfUndefined{supsetneqq}{\let\psup\@psup}{\let\psup=\supsetneqq}
	\def\@psub{\mathop{\mathop\subset\limits_{\neq}}}
	\IfUndefined{subsetneqq}{\let\psub\@psub}{\let\psub=\subsetneqq}
	\IfUndefined{preccurlyeq}{}{}
	\IfUndefined{npreceq}{}{}
	\mydefloglike{sgn}{sgn}\nolimits % sign

	\let\oldlor=\lor
	\let\oldland=\land
	\renewcommand\lor{\mathrel\oldlor}
	\renewcommand\land{\mathrel\oldland}

	\newcommand{\inv}[1]{{#1}^{-1}}
	\def\set#1for#2\eset{\left\{\:#1\mathbin\setmidchar#2\:\right\}}
	\let\setmidchar:
	\mydefloglike{dom}{domain}\nolimits
	\mydefloglike{supp}{supp}\nolimits % support (of a function)

	\let\sm\setminus \let\es\emptyset
	\newcommand\lp{\ifmmode\ell^p\else$\ell^p$\fi}
	\newcommand{\net}[3]{\ifmmode\set#1_{#2} for #2\in#3\eset
		\else$\net{#1}{#2}{#3}$\fi}

	\mydefloglike{relint}{relint}\nolimits
	\mydefloglike{co}{co}\nolimits % convex hull
	\mydefloglike{clco}{\overline{co}}\nolimits % closed convex hull
	\mydefloglike{diam}{diam}\nolimits % diameter of a set
	\mydefloglike{bdy}{bdy}\nolimits

	\mydefloglike{cor}{cor}\nolimits % core of a set
	\mydefloglike{Im}{Im}\nolimits % Image
	\mydefloglike{im}{im}\nolimits % image (alternative)
	\mydefloglike{Ker}{Ker}\nolimits % Kernel
	\mydefloglike{ker}{ker}\nolimits % kernel (alternative)
	\mydefloglike{rank}{rank}\nolimits % rank (dimension of image, usually)
	\mydefloglike{nullity}{nullity}\nolimits % nullity (dimension
	\mydefloglike{ch}{ch}\nolimits % characteristic poly
	\mydefloglike{diag}{diag}\nolimits % diagonal matrix
	\mydefloglike{lcm}{lcm}\nolimits % least common multiple
	\mydefloglike{lin}{lin}\nolimits % linear space spanned by
	\mydefloglike{tr}{tr}\nolimits % trace
	\mydefloglike{sp}{sp}\nolimits % linear space spanned by
	\mydefloglike{adj}{adj}\nolimits % adjoint
	\mydefloglike{gp}{gp}\nolimits % group generated by
	\mydefloglike{id}{id}\nolimits % ideal generated by
	\mydefloglike{charac}{char}\nolimits % characteristic
	\mydefloglike{Coker}{Coker}\nolimits % Cokernel
	\mydefloglike{Coim}{Coim}\nolimits % Coimage
	\mydefloglike{Ext}{Ext}\nolimits % Ext
	\mydefloglike{Tor}{Tor}\nolimits % Tor
	\mydefloglike{Hom}{Hom}\nolimits % Hom (ie. set of all homomorphisms)
	\mydefloglike{End}{End}\nolimits % End (ie. set of all endomormphisms)
	\mydefloglike{Inn}{Inn}\nolimits % Inn (ie. set inner automorphisms)
	\mydefloglike{Aut}{Aut}\nolimits % Aut (ie. set automorphisms)
	\mydefloglike{GF}{GF}\nolimits % finite fields
	\mydefloglike{charac}{char}\nolimits % characteristic
	\mydefloglike{Spec}{Spec}\nolimits % Spectrum of a ring
	\mydefloglike{Max}{Max}\nolimits % Maximal Spectrum of a ring
	\mydefloglike{Rad}{Rad}\nolimits % Radical of an ideal
	\renewcommand{\pmod}[1]{\allowbreak
		\ifinner\mkern9mu\else\mkern18mu\fi({\hbox{mod}}\,\,#1)}
	\mydefloglike{STS}{STS}\nolimits % Steiner triple system
	\mydefloglike{TTS}{TTS}\nolimits % twofold triple system
	\mydefloglike{DTS}{DTS}\nolimits % directed triple system
	\mydefloglike{BIBD}{BIBD}\nolimits % balanced incomplete binary design
	\mydefloglike{SBIBD}{SBIBD}\nolimits
	\mydefloglike{PBIBD}{PBIBD}\nolimits
	\mydefloglike{grad}{grad}\nolimits % gradient operator
	\let\del=\nabla
	\mydefloglike{div}{div}\nolimits % divergence operator
	\mydefloglike{curl}{curl}\nolimits % circurlation operator
	\mydefloglike{Int}{Int}\nolimits % interior
	\mydefloglike{inte}{int}\nolimits % interior (alternative)
	\mydefloglike{erf}{erf}\nolimits % error function
	\mydefloglike{erfc}{erfc}\nolimits % complementary error function
	\mydefloglike{rel}{rel}\nolimits % rel
	
	\mydefloglike{GL}{GL}\nolimits
	\mydefloglike{SL}{SL}\nolimits
	\mydefloglike{Sp}{Sp}\nolimits
	\mydefloglike{gl}{gl}\nolimits
	\let\@oldsl\sl
	\def\@newsl{\ifmmode\mathop{\operator@font sl}\nolimits\else\@oldsl\fi}
	\def\sl{\protect\@newsl}
	\mydefloglike{sp}{sp}\nolimits
	\let\te\otimes % tensor product
	\mydefloglike{Symm}{Symm}\nolimits % an alternative
	\renewcommand\o{^\bgroup\c@rc}
	\def\c@rc{\circ\futurelet\next\c@rcs}
	\def\c@rcs{\ifx*\next\let\nxt\c@rc@s \else\ifx^\next\let\nxt\c@rc@t
		\else\let\nxt\egroup\fi\fi \nxt}
	\def\c@rc@s#1{\c@rc} \def\c@rc@t#1#2{#2\egroup}

	\newcommand\DiracNotation{% changes meaning of < and > in math mode
		\mathcode`\<="8000 {\catcode`\<=\active \gdef<{{\delimiter"426830A}}}
		\mathcode`\>="8000 {\catcode`\>=\active \gdef>{{\delimiter"526930B}}}
		\mathcode`\@="8000 {\catcode`\@=\active \gdef@{{\dagger}}}
	}
	\mydefloglike{Re}{Re}\nolimits % real part
	\mydefloglike{Im}{Im}\nolimits % imag. part
	\mydefloglike{Arg}{Arg}\nolimits % principal argument
	\mydefloglike{Log}{Log}\nolimits % principal log
	\mydefloglike{res}{res}\nolimits % residue
	\mydefloglike{ord}{ord}\nolimits % order 
	\mydefloglike{arg}{arg}\nolimits
	\mydefloglike{cosec}{cosec}\nolimits
	\mydefloglike{edge}{edge}\nolimits % edge of an achronal set
	\mydefloglike{Lor}{Lor}\nolimits % Lorentz metrics
	\mydefloglike{PsR}{PsR}\nolimits % Pseudo-Riemannian metrics

\let\overbar\overline

\newcommand\Mhat{{\widehat{\cM}}}

\def\({\left(}
\def\){\right)}

\newcommand{\thisenumparens}[1]{%
	\@namedef{the\@enumctr}{{\rm(\csname#1\endcsname{\@enumctr})}}%
	\@namedef{label\@enumctr}{{\rm\csname the\@enumctr\endcsname}}%
	}
\newcommand{\enumparens}[2]{%
	\@namedef{theenum#1}{{\rm(\csname #2\endcsname{enum#1})}}%
	\@namedef{labelenum#1}{{\rm\csname theenum#1\endcsname}}%
	}
\newcommand{\thisenumparen}[1]{%
	\@namedef{the\@enumctr}{{\rm\csname #1\endcsname{\@enumctr}}}%
	\@namedef{label\@enumctr}{{\rm\csname the\@enumctr\endcsname)}}%
	}
\newcommand{\enumparen}[2]{%
	\@namedef{theenum#1}{{\rm\csname #2\endcsname{enum#1}}}%
	\@namedef{labelenum#1}{{\rm\csname theenum#1\endcsname)}}%
	}
\newcommand{\thisenumdot}[1]{%
	\@namedef{the\@enumctr}{{\rm\csname #1\endcsname{\@enumctr}}}%
	\@namedef{label\@enumctr}{{\rm\csname the\@enumctr\endcsname.}}%
	}
\newcommand{\enumdot}[2]{%
	\@namedef{theenum#1}{{\rm\csname #2\endcsname{enum#1}}}%
	\@namedef{labelenum#1}{{\rm\csname theenum#1\endcsname.}}%
	}
\enumdot i{arabic}
\enumparens{ii}{roman}
\enumparen{iii}{alph}

\def\p@enumi{}
\def\p@enumii{}
\def\p@enumiii{}
\def\p@enumiv{}

\newcommand{\deflabel}[1]{\label{def:#1}}
\newcommand{\thmlabel}[1]{\label{thm:#1}}

\newcommand{\figlabel}[1]{\label{fig:#1}}

\newcommand{\seclabel}[1]{\label{sec:#1}}
\newcommand{\enlabel}[1]{\label{en:#1}}
\newcommand{\eqlabel}[1]{\label{eq:#1}}

\newcommand{\defref}[1]{\ref{def:#1}}
\newcommand{\figref}[1]{\ref{fig:#1}}

\newcommand{\thmref}[1]{\ref{thm:#1}}
\newcommand{\secref}[1]{\ref{sec:#1}}
\newcommand{\enref}[1]{{\rm\ref{en:#1}}}
\@ifundefined{eqref}{% i.e. no amsmath package
	\newcommand{\eqref}[1]{{\rm(\ref{eq:#1})}}
}{
	\let\ams@eqref\eqref
	\renewcommand{\eqref}[1]{{\ams@eqref{eq:#1}}}
}

\newcommand{\underuparrow}[1]{\mathop{\vtop{\ialign{##\crcr
  $\hfil\displaystyle\strut{#1}\hfil$\crcr\noalign{\kern3\p@\nointerlineskip}
  $\hfil\uparrow\hfil$\crcr\noalign{\kern3\p@}}}}\limits}
\newcommand{\underaccent}[2]{\smash{%
	\vtop{\ialign{\hfil##\hfil\crcr
	$\m@th#2{}$\crcr\noalign{\kern2pt\nointerlineskip}%
	\vtop to1pt{\hbox{$\m@th#1{\phantom{x}}$}\vss}\crcr
	\noalign{\kern1pt\nointerlineskip}}}}}

\newcommand{\circover}[1]{\vbox{\ialign{\hfil##\hfil\crcr
	$\m@th\scriptstyle\circ$\crcr\noalign{\kern1pt\nointerlineskip}
	$\m@th#1$\crcr}}}

\let\oldvec=\vec % the old \vec puts an arrow over the thing, yuk.
\renewcommand\vec[1]{{\bold{#1}}}
\IfUndefined{bold}{\let\bold\oldvec\let\vec\oldvec}{}
\mathcode`\^^V="8000
{\catcode`\^^V=\active\global\def^^V{\vec}} % expansion of \vec deferred

\IfUndefined{geqslant}{}{%
	\let\ge\geqslant
	\let\geq\ge
	\let\le\leqslant
	\let\leq\le
}

\IfUndefined{leqneqq}{}{}

\IfUndefined{varepsilon}{}{\let\epsilon\varepsilon}

\newcommand{\range}[3]{\ifmmode
		#1_{#2},\ldots,#1_{#3}
	\else
		$#1_{#2},\ldots,#1_{#3}$%
	\fi}

\IfUndefined{mathcal}{\let\mathcal\cal}{}

\newcommand\cA{{\mathcal A}} \newcommand\cB{{\mathcal B}}
 
\newcommand\cE{{\mathcal E}}

 \newcommand\cL{{\mathcal L}}

 \newcommand\cR{{\mathcal R}}
 
\newcommand\cU{{\mathcal U}} 
\newcommand\cW{{\mathcal W}}

\let\b=\beta

\let\vp=\varphi  % a shorthand for \varphi - NOT \phi (anymore)
\let\e=\epsilon

\let\d=\partial %%%%%% <------- NOTE!
\let\de=\delta

\let\g=\gamma
\let\G=\Gamma

\let\L=\Lambda

\let\a=\alpha

\let\w=\omega

\IfUndefined{mathrsfs}{}{}
\IfUndefined{mathbb}%
	{\IfUndefined{Bbb}{\global\let\mathbb\bf}{\global\let\mathbb\Bbb}}{}

\newcommand\R{{\mathbb R}}

\mathcode`\*="8000
{\catcode`*=\active \gdef*{^\bgroup\st@r}}
\def\st@r{\ast\futurelet\next\st@rs}
\def\st@rs{\ifx*\next\let\nxt\st@@@s \else\ifx^\next\let\nxt\st@@@t
	\else\let\nxt\egroup\fi\fi \nxt}
\def\st@@@s#1{\st@r} \def\st@@@t#1#2{#2\egroup}

\IfUndefined{citeasnoun}{%
	\let\citeasnoun\cite
}{%
}

\IfUndefined{rom}{%
	\IfUndefined{XSselectfont}{%
		\newcommand{\rom}[1]{{\rm#1}}
	}{%
		\let\rom\relax
	}
}

\IfUndefined{looparrowright}{}{} % if no amsfonts

\makeatother

\mydefloglike{gptprt}{gptprt}\nolimits

\def\lto#1#2{\stackrel{#1\to\infty}\longrightarrow #2 }

\mydefloglike{essinf}{ess\,inf}\nolimits
\mydefloglike{esssup}{ess\,sup}\nolimits
\mydefloglike{essrange}{ess\,range}\nolimits

\mydefloglike{epi}{epi}\nolimits

\def\lvert{\mathopen|} \def\rvert{\mathclose|}

\def\Mhat{\widehat M}
\def\deM{\d_{\cE}M}
\def\pMb{{\overbar{\phi(M)}}}
\newcommand{\LL}{{\boldsymbol\L}}

\begin{document}

\title[A rigidity result for smooth ideal boundaries]{A rigidity
result on the ideal boundary structure of smooth space-times}
\author{C. J. Fama\dag\footnote[3]{E-mail: {\tt chris@maths.uq.edu.au}.}
 and C. J. S. Clarke\ddag\footnote[4]{E-mail: {\tt
cjsc@maths.soton.ac.uk}.}}
\address{\dag\ Department of
Mathematics, The University of Queensland, Brisbane 4072, 
     Australia.}
\address{\ddag\ Faculty of Mathematical Studies,
 University of Southampton,
 Southampton SO17 1BJ, UK.}
\date{November 1997}
\jl{6} % C&QG
\submitted
%\maketitle

\begin{abstract}
Following a survey of the abstract boundary definition of Scott and
Szekeres, a rigidity result is proved for the smooth case, showing that
the topological structure of the regular part of this boundary in
invariantly defined.
\end{abstract}

\section{Introduction}

The most intuitively natural definition of a boundary to a space-time
$(M,g)$ would seem to be obtained by topologically embedding $M$
in a larger manifold $\Mhat$ {\em of the same dimension\/} by means of
a map $\phi:M\to\Mhat$, followed by taking the topological boundary
$\d_\phi M\defeq\d(\phi(M))$ of the image $\phi(M)$ in
$\Mhat$. (Where $\phi$ is of differentiability class $C^\ell$,
$\ell\ge1$ we shall refer to this as a $C^\ell$ {\em envelopment\/},
or simply a {\em smooth envelopment} if the class $C^\ell$ is
understood.) The problem with this is that, unless further conditions
are imposed, $\d_\phi M$ depends critically on the choice of $\phi$,
the different possible $\d_\phi M$ for different $\phi$ being so
variable in structure that they contain no invariant information.

Because of this difficulty, this approach was abandoned early on the
history of relativity, in favour of boundary constructions linked more
closely to the detailed geometry of $M$. If, however, we were to
restrict ourselve to {\em regular envelopments\/}, meaning
envelopments where $\phi$ was a metric isometry into a ``regular'' (in
some sense) $(\Mhat, \hat g)$, then we might hope for more rigidity
(that is, constancy of topological structure) in $\d_\phi M$. To
provide an overall context for this and related constructions Scott
and Szekeres \cite{Scott94} have developed the {\em abstract boundary
construction\/} which incorporates all possible boundaries $\d_\phi M$
with appropriate identifications being made between corresponding
subsets of the boundaries (boundary sets).  It is already known that
if the metric is used certain topological properties of boundary sets
{\em are\/} preserved under equivalence, even allowing for quite
unpleasant properties of the metric \cite{Fama95}.  This work cited
can be said to serve as an interesting example of the relative ease
with which certain simple results about the abstract boundary can be
obtained, and also the crucial r\^ole played by our
``regularity/extendibility'' assumptions (q.v.); however, the work is
not used here.

In this paper we review the abstract boundary construction, we
survey possible definitions of the notion of ``regular'', and we
show within this framework that we can achieve an appropriate rigidity
of structure for a boudary constructed from regular envelopments.
More precisely, we show that if we consider only boundaries that are
regular and satisfy a Lipschitz condition, then all representatives of
an equivalence class of boundary sets, in the sense of the abstract
boundary, are homeomorphic.

In what follows, by a {\em pseudo-Riemannian manifold\/} $(M,g)$ we will
mean a Hausdorff, paracompact topological manifold $M$, without a
boundary in the usual sense, that is equipped with a $C^1$ atlas and a
metric $g$.%
\footnote{
	Note that we require all manifolds to have differentiable
	structure, but make no demands on $g$ at the moment, beyond
	continuity; note also that saying that $(M,g)$ is $C^k$ means that $g$
	is $C^k$, and the atlas of $M$ is at least $C^{k+1}$.
}
If, further, $M$ is connected, $n=\dim M\ge2$, and the metric $g$
is of Lorentz signature $(\mathord- \mathord+ \cdots \mathord+)$, then
we shall refer to $(M,g)$ as a {\em space-time}.  If $g$ is
$C^\ell$, some $\ell\ge1$, the
covariant differential with respect to the unique torsion-free
metric-compatible connection on $M$ will be denoted $\del$, or
$\stackrel g\del$ if we wish to emphasise the r\^ole of the metric.
In fact, we mostly use the notation of \citeasnoun{O'Neill83}.  For
example, $X_p \in T_p(M) $ will denote the value of the vector
field $X$ at the point $p\in M$.

In referring to envelopments a {\em boundary set\/}  is a subset
$B\sub\d_\phi$.  (We will normally have primed objects belonging to an
envelopment $(M,\Mhat',\phi')$, with corresponding unprimed objects
belonging to an envelopment $(M,\Mhat,\phi)$.)

\section{Regular abstract boundaries}
\seclabel{regular abstract boundary}

The equivalence relation used in the definition of the abstract boundary
is as follows.  Suppose that we are
given boundary sets $B,\,B'$ of two envelopments
$(M,\Mhat,\phi)$, $(M,\Mhat',\phi')$, respectively. We say that
$B$ {\em covers\/} $B'$ ($B\rhd B'$)
if for every open neighbourhood $\cW$ of $B$ in $\Mhat$, there is an open
neighbourhood $\cW'$ of $B'$ in $\Mhat'$ such that
\begin{equation}
	\eqlabel{covering, original}
	\phi\circ(\phi'\inv)(\cW'\cap\phi'(M)) \sub \cW.
\end{equation}
It is easily seen that $B\rhd B'$ if and only if
``one cannot approach $B'$ from within $M$ without also approaching $B$''.
In a sense, then, this means that $B$ is ``bigger'' than $B'$, and this
is the reason for the notation.
As an example, let
\begin{eqnarray}
	\Mhat=\Mhat'=\R^n,\\
	M=\R^n\sm\{\mathbf0\}\approx \R\times S^{n-1},\\
	\phi={}\hbox{inclusion},\\
	\phi'(r,\Theta)=(r+1,\Theta)\in\R^n\sm\{0\}.
\end{eqnarray}
Then $p=\mathbf0\in\d_\phi$ covers {\em any\/} subset of $\d_{\phi'}$.

One says that boundary sets $B$ and $B'$ are {\em equivalent},
$B\sim B'$, if they cover each other: $B\rhd B'$ and $B\lhd B'$.
The equivalence class containing the boundary set $B$ is denoted $[B]$,
and is called an {\em abstract boundary set}.
If $[B]$ contains a singleton boundary set $\{p\}$, then $[B]$ may
also be denoted by $[p]$, and is called an {\em abstract boundary point\/}.%

The collection of all abstract boundary points constitutes
the {\em abstract boundary\/} $\cB M$ of $M$; that is,
\[
	\cB(M) \defeq
	\set [p] for p\in\d_\phi M \hbox{ for some envelopment }
		(M,\Mhat,\phi) \eset.
\]

According to this definition, boundary points and abstract boundary
points admit quite a rich further classification \cite{Scott94}, which
need not concern us here.  Note that the abstract boundary
construction is not as developed (or, perhaps, useful!) as such
constructions as the bundle boundary ($b$-boundary) of Schmidt
\cite{Schmidt71,Clarke93b,Dodson78,Dodson80}, the conformal boundary
($c$-boundary) of Geroch {\em et al.}
\cite{Geroch72,Beem81,Clarke93b}, or even the ``$A$-boundary'' of
Clarke \cite{Clarke93b}.  The latter turns out to be very closely
related to a certain (``Lipschitz'') {\em regular abstract boundary}.

We now introduce the r\^ole of a metric.
Let $(M,\Mhat,\phi)$ be an envelopment of a $C^k,\,k\ge1$
pseudo-Riemannian manifold $(M,g)$. A point $p\in\d_\phi$, which has
a neighbourhood $\cU_p$ such that $\Mhat_p\defeq\phi(M)\cup\cU_p$
can be endowed with a $C^\ell$ (some $1\le \ell\le k$)
pseudo-Riemannian metric $\hat g$ extending $g$, i.e.,
$g=\psi^\ast\hat g$, is called $C^\ell$-{\em regular}.
The envelopment may then be referred to as $(M,g,\Mhat,\hat g, \phi)$.

For regular points we will use both the original setting of the
abstract boundary \cite{Scott94} just described, and also a definition
from earlier, unpublished work of C.~J.~S.~Clarke and S.~M.~Scott on
defining topologies on the abstract boundary.  (Actually, the authors
of \citeasnoun{Scott94} mention that their classification requires
merely an affine connection, not necessarily the Levi-Civit\`a
connection of a pseudo-Riemannian metric.)

\begin{definition}
	Let $\cE$ be a collection of envelopments and
	\[
		X_\cE \defeq \sum_{\phi\in\cE}\pMb
		\qquad\hbox{(a disjoint union)},
	\]
	and define a relation $\sim$ on $X_\cE$ in the following way:
	\begin{enumerate}\thisenumparens{roman}
	\item
		If $x\in\d_\phi$, $y\in\d_\psi$,
		then $\sim$ is the usual boundary set equivalence.
	\item
		If $x=\phi(\xi)$ with $\xi\in M$, then $y\sim x\iff y=\psi(\xi)$,
		some $\psi\in\cE$.
	\end{enumerate}
	This is an equivalence relation, and we set $M_{\cE}=X_\cE/\mathord\sim$.
	Let $\pi$ be the projection $X_\cE\to M_{\cE}$, identify
	\[
		 M \qquad\hbox{with} \qquad
		\set [\phi(p)] for p\in M,\,\phi\in\cE\eset,
	\]
	and write $\deM$ (or $\d_\cE$) for $M_{\cE}\sm M$.

	We will use the notation $\phi:M\to\Mhat_{\phi}$ for envelopments,
	and for later use will also assume, as is entirely reasonable,
	that each $\Mhat_\phi$ is assumed to have an atlas
	$\cA_{\Mhat_{\phi}}$ for which the ``pullback''
	$\phi*\cA_{\Mhat_\phi}$ is equivalent to the atlas $\cA_M$ on $M$.
	(We will impose a further condition on atlases in \S\secref{lip}.)
\end{definition}

The key use of the above formalism is that it allows us to vary
construction of the completion $M_{\cE}$ as we vary our chosen set of
envelopments $\cE$. The usual abstract boundary is $\d_{\cE_0} M$,
where $\cE_0$ is the collection of {\em all\/} smooth envelopments
$\phi$ of $ M$ into some smooth manifold $\Mhat_\phi$. (In
order to be absolutely correct mathematically, ``all'' needs to be
qualified in such a way as to ensure that $\cE$ is a set and not a
proper class---for instance by working in a category of concretely
defined manifolds.)

In the next section we will survey in detail possible candidates for
``regular'' maps, metrics etc.\ in terms of a class $\L$ of maps
between Euclidean spaces. For a given choice of $\L$ we will denote by
${\LL}$ the corresponding category of class $\L$ maps between
manifolds {(e.g.${\LL}={\mathbf C^k}$),} and by $\L(M.N)$ the set
of morphisms of ${\LL}$ from manifold $M$ to manifold $N$.

By, e.g., the categorical statement``$\LL\sub\mathbf
C^1$'' we mean that $\L(M,N)\sub C^1(M,N)$ for all $M,N$ under
consideration.

Armed with a suitable category of maps, then, we make an important 
set of definitions, the first of which generalises our earlier
notion of ``regularity'' {for $ {\LL}\supsetneq {\mathbf
C^1}$, but is identical to the earlier notion for
$ {\LL}=\mathbf C^1$}.

\begin{definition}
\deflabel{extendibility...}
	Let $\LL$ be a category of maps as above, and           
	\[
		g\in\L(M,\PsR^n_\nu(M))\cap\G(M,\PsR^n_\nu(M))
	\]
	($\G(B,T)$ denotes the set of sections $B\to T$) be a given
	pseudo-Riemannian metric of signature $(n,n-\nu)$, {\em i.e.}, a
	non-degenerate, symmetric bilinear form at every point of $M$.  We
	shall say that $g$ is {\em $\L$-extendible about $p$}, where
	$p\in\d_\phi$ for some envelopment $\phi\in\cE_0$, if the
	following holds: there is a neighbourhood $U_p$ of $p$ (in
	$\Mhat_\phi$) and a pseudo-Riemannian metric
	$\hat g_p$ on $U_p$, $\hat
	g_p\in\L(U_p,\PsR^n_\nu(U_p))$, for which $\phi^\ast\hat
	g_p=g$. Here $\PsR^n_\nu(M)$ denotes the subbundle of the
	symmetric tensor product $\Symm(T*M\te T*M)$ with fibres
	consisting of non-degenerate bilinear forms of the appropriate
	signature.  Now let
	\begin{eqnarray*}
		\cE[\LL] = \set \phi\in\cE_0 for \forall
			p\in\d_\phi M,\,g \hbox{ is $\L$-extendible about } p\eset,\\
		\cR[\LL](M) = \set [p]_\sim\in\d_{\cE_0}M for g \hbox{ is
			$\L$-extendible about } p, \hbox{ where }p\in\d_\phi\eset.
	\end{eqnarray*}
	We call this a {\em regular abstract boundary}.
\end{definition}
Note that $\cR[\LL]$ can be identified, loosely at least, with the
abstract boundary $\d_{\cE[\LL]} M$ associated with the collection of
all envelopments with ``everywhere $\L$-regular'' boundaries.

One would expect these to be useful abstract boundaries in relativity
theories, where we would like to seek properties of {\em all\/}
possible metric extensions (of some class) through a ``boundary
point'', without using or requiring detailed knowledge of the analytic
form of the metric involved.

\section{Various regularity classes}

We mention a few possible choices for our extension classes
$\L(\spdot)$, although we will only consider the first in this paper.
For clarity we assume that we are speaking of Lorentzian metrics.

\begin{description}
\item[``Smooth regularity'',
 \boldmath$\LL=\mathbf C^k,\,k\ge1$.\unboldmath]
	Such regular abstract boundaries, the first to be considered, are
	the subject of unpublished work by C.~J.~S.~Clarke and S.~M.~Scott.
\item[``Existence theorem regularity.'']
	There are several possibilities that suggest themselves
	immediately here, relying on the existence theorems of
	\citeasnoun{Hughes76}.  These rely on conditions such as the
	difference between the metric that we consider and a fixed
	``asymptotically flat'' or ``background'' metric lying in the
	Sobolev space $H^{2.5+\e}(\R^{n-1}),\,\e>0$ (the metrics checked
	are the restrictions of a full space-time metric to hypersurfaces).
	We could allow extensions such that through any point there is a
	spacelike hypersurface on which induced data satisfies the
	conditions; or such that through any point there is a set of
	space-like hypersurfaces whose normals at the point form an open
	set and which satisfy the conditions, or such that there is a
	(local or global) foliation satisfying the conditions.
\item[``Non-quantum physics regularity.'']
		It could be argued that a fairly robust criterion for pair
		production to become significant within a 3-dimensional region
		of ``size'' $L$ where the components of the Riemann tensor in
		a ``reasonable'' frame are larger than $R$ is that
		\[
			R > \max(m^2c^2/\hbar^2, L^{-2})
		\]
		the argument being either merely on dimensional grounds, or by
		requiring that virtual pairs acquire sufficient energy within
		the region during the Heisenberg uncertainty time to become
		non-virtual.  

	We would, then, require $\L$ to consist of metrics for which this
	condition did {\em not\/} hold, i.e., those metrics where we might
	expect Einstein's theory to describe ``reality'' accurately.
\item[``Distributional (Cauchy-Schwarz or Colombeau) regularity.'']
	Here we would, technically, go outside our formalism and allow
	extensions---and presumably, although not necessarily, interior
	metrics also---that were not functions at all, but either
	Cauchy-Schwarz (linear) distributions, or Colombeau's
	``generalised functions'' (non-linear distributions).  To cope
	with such generality, we would probably have to impose further
	cnditions on the Levi-Civit\`a connection components of these
	extended metrics, e.g., square-integrability in the first case and
	integrability in the second.  Physically, these would describe,
	respectively, impulsive gravitational waves in their most obvious
	form, and ``stringy'' space-times admitting conical singularities.
\item[``Geroch--Traschen regularity'' \cite{Geroch87},]
	{$\LL=\mathbf C^0\cap\mathbf W^{1,2}_{loc}$}  (intersection of
	categories of maps being defined by the intersection of function classes)
	
	We call these {\em metrics of Geroch--Traschen
	type}. (and the second.....) deleted  Physically, these appear to
	encompass gravitational wave space-times, though the metric must
	be transformed so that it is continuous.
	This ``Rosen form'' transformation typically destroys at least the
	$C^1$ structure of the manifold in question.  We will propose a
	way to deal with this in a forthcoming publication (in which we
	will also allow degenerate metrics, though still requiring
	square-integrability of Levi-Civit\`a connection components).
\end{description}

In this paper we consider only the first of these items.

\section{Lipschitz boundaries}
\seclabel{lip}

In this section we recall some material from \citeasnoun{Clarke93b},
in which certain proscriptions on the nature of an atlas entail that
the {\em A-boundary} can be constructed.  This turns out to be a
rather useful boundary construction for later work of Clarke
and colleagues on singularity theorems, appearing in the same
reference.  It amounts to the    requirement that the boundary ``not
wiggle about too much''\ldots

Let $\cA=\{(W_\a,p_\a)\}_{\a\in A}$ be an atlas for an envelopment
$\Mhat_\phi$ of $M$, and write $U_\a=p_\a(W_\a\cap \phi(M))\sub
p_\a(W_\a)\sub\R^n$.  For brevity, we will write $ U_\a*$ for that
part of the boundary of the open set $U_\a$, which, speaking somewhat
loosely, corresponds to parts of ``the boundary of $M$'' which ``have
two sides'' and are ``regular'' with respect to our chosen $g$ and
$\L$:
\begin{eqnarray}
 \fl 	U_\a*[g,\L] \defeq\set  x\in
		p_\a(\d(\phi(M))\cap W_\a) for g \hbox{ is
      $\L$-extendible about }\inv{(p_\a)}(x)\eset \nonumber\\
	 \qquad{} \sm \set x\in\d U_\a for \hbox{a neighbourhood of
		$x$ is contained in } \overbar{ U_\a} \eset, \eqlabel{U* def}
\end{eqnarray}
See Figure~\figref{lip boundary}.
\begin{figure}
	\centering
	\includegraphics[angle=0,width=.8\textwidth]{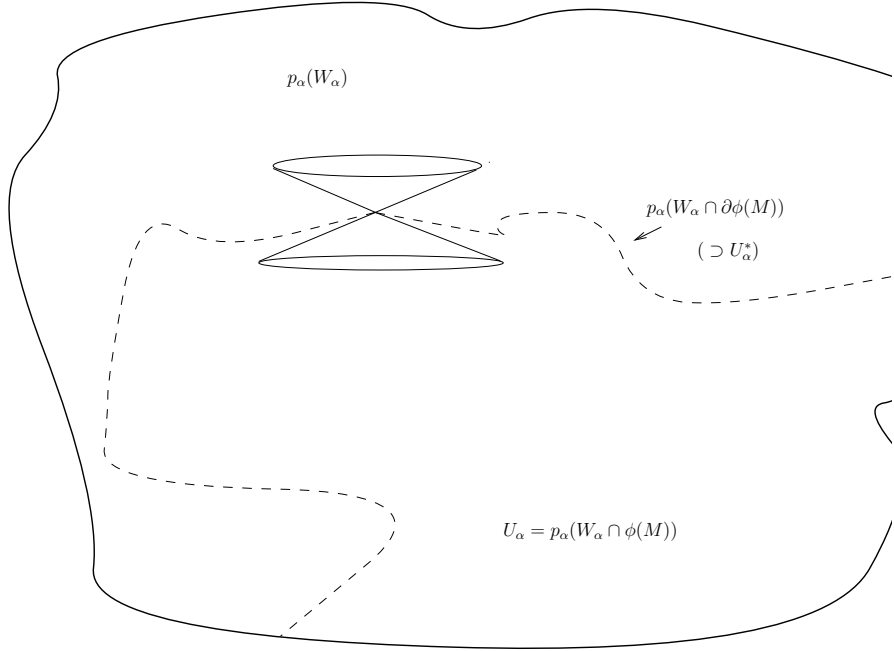}
	\caption{A typical Lipschitz boundary, on which a pair of
		proper (inward and outward) cones are shown.\figlabel{lip boundary}}
\end{figure}

Note that we exclude points of $\inte\bar U_\a$ from the definition of
$U_\a*$.  This is because portions of a boundary which are not
``Lipschitz hypersurfaces'' cannot make up the ``Lipschitz
boundaries'' that we define, following \citeasnoun{Clarke93b}, in a
moment. (In particular, this excludes higher codimension potential
boundaries like the $z$-axis for $M=U=\R^3\sm\{z$-axis$\}$,
$\phi={}$inclusion.)

Finally, we impose a regularity condition on our chosen points
$U_\a$.

\begin{definition}
\deflabel{lip boundary}
	We say that the envelopment $\Mhat_\phi$ has a {\em Lipschitz
	boundary\/} if the following three conditions hold, for each of
	the charts $W_\a$ (the last condition, depending on {\em pairs\/}
	of charts, is for the benefit of the A-boundary construction and
	will not concern us for the remainder of this work):
	\begin{enumerate}
	\def\labelenumi{(\Alph{enumi}${}_\a$)}
	\let\theenumi\labelenumi
	\item\enlabel{lip boundary 1}
		$\bar U_\a$ is compact in $\R^n$.
	\item\enlabel{lip boundary 2}
		There is a vector field $k_\alpha:\R^n\to\R^n$ (assumed to be
		as smooth as the atlases $\cA$, $\cA_\phi$), such that for
		each $x\in U_\a*$, we have $\lvert k_\alpha(x)\rvert=1$ and
		the existence of some $\de=\de(x)>0$ with
		\[
			y\in U_\a\iff k_\alpha(x)\cdot(y-x)<f_x(P_{k_\alpha(x)}(y-x))
		\]
		if $\enorm{y-x}<\de$.
		Here $\enorm\spdot$ denotes the Euclidean norm, $f_x$ is some
		Lipschitz function $\R^n\to \R$, and $P_vy;y\mapsto y-(v\cdot y)v$ is
		the orthogonal projection from $\R^n$ onto $v^\perp$.
	\def\labelenumi{(\Alph{enumi}${}_{\a\b}$)}
	\let\theenumi\labelenumi
	\item\enlabel{lip boundary 3}
		Writing $\psi_{\a\b}\defeq =p_\a\circ \inv{p_\b}$ for the
		transition functions of $\cA_\phi$, $\lvert
		D\psi_{\a\b}\rvert$ is bounded on $U_\a\cap U_\b$.
	\end{enumerate}
\end{definition}

Note that the choice $\phi={}$identity, $\cA=\cA_\phi$ leads to a case
of the A-boundary construction, where the subsets $U_i*$ of $\d U_i$
are taken to be the maximal such subsets through which $g$ is
$\L$-extendible.  (The union of $M$ with the A-boundary $\bar M\sm M$
constructed in \citeasnoun{Clarke93b} can be exhibited as a ($C^0$)
manifold-with-boundary $\bar M\defeq\bar M^{(A)}$ (the details are in
the quoted reference).)  We shall call this the {\em maximally regular
\rom(Clarke\rom) A-boundary\/} construction.

We examine condition \enref{lip boundary 2}, the core of the matter, a
little more closely.  If $A:\R^n\to \R^n$ is a non-singular homothety
(i.e., the composition of a dilation, translation, and rotation) with
$A_\ast$ taking the unit vector $k_\alpha(x)$ to $(0,\ldots,0,1)$ and
$A(x)=0\in\R^n$, then this is just the condition that
\[
	U_\a\cap B(0,\de)=\inv A\set z\in\R^n for \enorm z<\de,\,
		z^n<\hat f_x(z^1,\ldots,z^{n-1})\eset
\]
for some Lipschitz $\hat f_x:\R^{n-1}\to \R$ with $\hat f_x(0)=0$.
It is clear, then, that this is equivalent to
\[
	U_\a*\cap B(0,\de)=\inv A\set z\in\R^n for \enorm z<\de,\,
		z^n=\hat f_x(z^1,\ldots,z^{n-1})\eset.
\]
Let $\hat K$ be a Lipschitz constant for $\hat f_x$; it follows that
\[
\fl \forall z\in U_\a*\cap B(x,\de(x)),\quad
	(z-\circover C)\cap B(x,\de)\sub U_\a\hbox{ and }
 	(z+\circover C)\cap B(x,\de)\sub \R^n\sm\bar U_\a,
\]
where $\circover C\defeq\R^{>0}\cdot [B_{\R^{n-1}}(0,1)
\times\{\hat K\}]$ is the interior of a cone.%

We give these cones $z\pm\circover C$ names: $z-\circover C$ (or
rather, $(\psi_\a\inv)\(\inv A(z-\circover C)\)$) we call
a {\em proper inward cone\/} and that with ``$+$'' rather than ``$-$''
a {\em proper outward cone}.

\begin{proposition}[Proposition 3.1, \citeasnoun{Clarke93b}]
\thmlabel{proper inward and outward cones}
	Condition \enref{lip boundary 2} of the definition of a Lipschitz
	boundary (Definition~\defref{lip boundary}) is equivalent to the
	existence of proper inward and outward cones everywhere near
	$U_\a*$.
\end{proposition}
\begin{remark}
	In the reference quoted there is no reference to (what we call)
	proper outward cones---but {\em if condition \enref{lip boundary
	2} holds}, there must be a proper outward cone as well as a proper
	inward cone.
\end{remark}

The existence of proper inward cones will be used in
the next section. First we make a new definition.

\begin{definition}
	Consider the class of envelopments
	\[
		\cL\cE[\LL]\defeq\set\phi\in\cE[\LL]  for \phi\hbox{ has a
			Lipschitz boundary}\eset,
	\]
	and $\cL\cR[\LL]\defeq\d_{\cL\cE[\LL]}$ (denoted $\cL\cR[M,g,\LL]$ if
	we want to highlight the manifold $M$ and metric $g$), the
	abstract boundary constructed from this set of envelopments.  We
	shall call the latter the {\em Lipschitz regular abstract
	boundary}. (Similary we shall call the appropriate equivalence classes
		{\em Lipschitz regular abstract boundary points}, and their
		representatives {\em Lipschitz regular boundary sets}, and so
		on.)

	Finally, we shall write $\cL\cR*[M,g,\L]$ for the class of all
	equivalents of boundary sets, without any demand that it contain a
	point as a representative.  (This makes some discussions slightly
	easier.)
\end{definition}

\section{The bundle metric---classical ($C^1$) case}
\seclabel{classical frame}

If $\LL\sub\mathbf C^1$ and $g\in\mathbf C^1$ then
there is a (continuous) connection on $(M,g)$ and the frame bundle
$LM$ (structure group $GL(n,\R)$) admits a topological metric $d$,
induced by a $C^0$ Riemannian structure as in
\citeasnoun[Chap.~3]{Clarke93b}, \citeasnoun{Dodson80} or
\citeasnoun{Dodson78}.

The same construction gives a topological metric $\hat d_p$ on
$L(\phi(M)\cup U_p)$ ($p\in \d_\phi$), resulting
from the (continuous) Levi-Civit\`a connection of $\hat g_p$.
However, there are at least two potential problems with this.

Firstly, the metric in the frame bundle depends on the boundary point:
$\hat d_p$ depends on the point $p$.  This, however, will not prove to
be a problem when we come to the crux of the matter,
Theorem~\thmref{need lip boundary}, but we must keep the subscript $p$
in mind!

\begin{figure}
	\centering
	\includegraphics[angle=0,width=.7\textwidth]{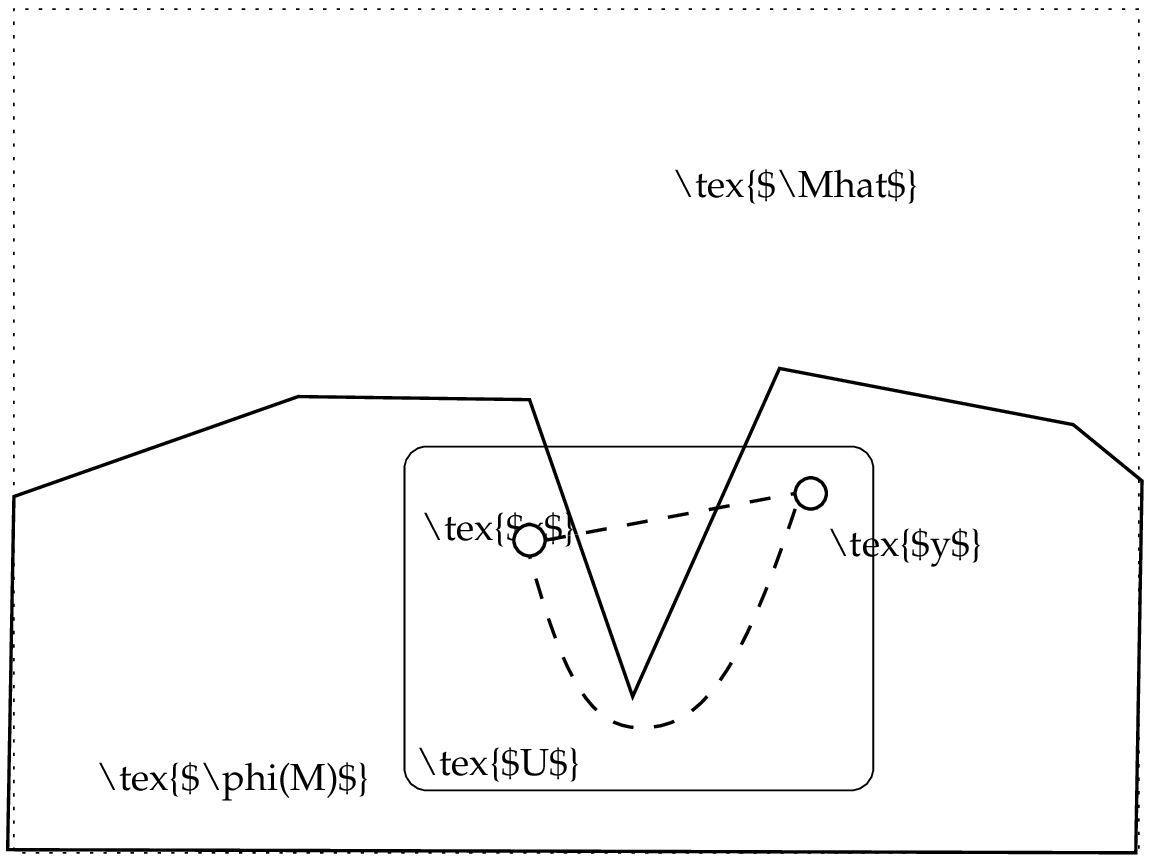}
	\caption{$\hat d(\tilde x,\tilde y)<d(\tilde x,\tilde y)$ because
		there is a curve from $x$ to $y$ in $\Mhat_\phi$ of strictly
		shorter g.a.p.\ length than any in $\phi(M)$.
		\figlabel{dhat-minimizer out of M}}
\end{figure}

Secondly, we have the following matter which rather complicates our
procedings.  For points $x,y\in\phi(M)$, we certainly have $\hat
d_p(\tilde x,\tilde y)\le d(\tilde x,\tilde y)$ (tildes denoting points
of fibres as usual)---loosely, $\hat d_p\le d$---but we may have $\hat
d_p(\tilde x,\tilde y)<d(\tilde x,\tilde y)$.  This is evident from
Figure~\figref{dhat-minimizer out of M} (take $\hat g_p$ to be flat).
Unfortunately, in a crucial argument below (Theorem~\thmref{need lip
boundary}) we will want to have $\hat d_p(\tilde x_i,\tilde
y_i)\to0\implies d(\tilde x_i,\tilde y_i)\to0$.  The inequality ``goes
the wrong way''!

\subsection{Using Lipschitz boundaries to imbed the Cauchy completion
$\overbar{LM}$ in $\overbar{L\Mhat}$}
\seclabel{after potential problems}

It is quite sufficient for our needs, however, to have a mere bound
``$\phi_\ast d \leq\mbox{constant}\times\hat d_p$'', in a small
neighbourhood of each point of the boundary.  We will call into play
the Lipschitz condition imposed on the boundary in the last section.
But first, a technical (but straightforward) lemma.  A sketched proof
only will be given---the reader unsatisfied with this is invited to
turn to the more rigourous, and general, treatment of a forthcoming
paper.  For now, though, note that we require that
$\LL\sub\mathbf C^1$.

\begin{lemma}
\thmlabel{hand-wave gap dist approxs Eucl}
	Let $g$ be a $C^1$ pseudo-Riemannian metric on $\R^n$.  Then,
	given a chart $(\widetilde U,\widetilde\vp)$, a point
	$y\in\widetilde U$, and any positive $\e>0$, we can find a small
	neighbourhood $U\sub\widetilde U$ of $y$ in which $\lvert
	\ell_1(\g)-\ell_2(\g)\rvert<\e$ for all curves $\g:[0,1]\to U$
	whose images under $\widetilde\vp$ are straight lines.  Here
	$\ell_1(\g)=k \times \int_{\g}\enorm{\mathbf\w(\dot\g)}$,
	$\mathbf\w=\{w^a\}$ being a parallel frame along $\g$, is a
	g.a.p.\ length, $k$ is a constant depending only on the frame at
	$\g(0)$, and $\ell_2(\g)=\int_\g\enorm{\dot\g}$ denotes Euclidean
	length.
\end{lemma}
\begin{proof}[sketch]
	Given $y$, we can choose $U$ so small that $\sup_{x\in
	U,i,j}\lvert g_{ij}((x)-\eta_{ij}\rvert$ and $\sup_{x\in
	U,i,j,k}\lvert g_{ij,k}((x)\rvert$ are as small as desired (i.e.,
	$g$ is close to a pseudo-Euclidean metric in the $C^1$ strong
	Stiefel-Whitney topology \cite{Hirsch76}), so the Christoffel
	symbols of the metric $g$ are as small as desired.  We use the
	usual continuous dependance of the solutions of a ($C^1$)
	differential equation on its parameters (see, e.g.,
	Theorem~IV.2.1~of \citeasnoun{Hartman64}), here specifying our
	curve by means of two extra parameters, namely its endpoints.  The
	solution parallelly transported vector field depends continuously
	on the (small) Christoffel symbols of $g$, so the solution is
	close to that for a flat metric, so for a short curve $\g$,
	\[
		\enorm{\boldsymbol\w(\dot\g)} \approx c \times\enorm{\dot\g}
	\]
	where $c $ depends only on $\boldsymbol\w_{\g(0)}$. (Of course it
	depends on the metric $g$, but this is fixed once and for all.)
	This ``closeness'' is uniform for ``endpoint parameters'' in some
	neighbourhood $U\sub\R^n\times\R^n$ of the origin, by the
	continuous dependence cited. We get the result.
\end{proof}

\begin{theorem}
\thmlabel{need lip boundary}
	Consider an envelopment $\phi\in\cE[\LL]$,
	$\LL\sub\mathbf C^1$, with a Lipschitz boundary, where
	$g\in\mathbf C^1$.

	Let $p\in \d_\phi$. There is a neighbourhood $U$ of $p$ \rom(in
	$\Mhat_\phi$) such that we cannot have sequences $\seq{\tilde
	x}i,\seq{\tilde y}i$ of $LM$, lying over points of $U$, for which
	$\hat d_p(\phi_\ast\tilde x_i,\phi_\ast\tilde y_i)\lto i0$ while
	$d(\tilde x_i,\tilde y_i)$ is bounded away from zero.
\end{theorem}
\begin{proof}
	Assume given sequences $\seq{\tilde x}i,\seq{\tilde y}i$ of $LM$
	for which $\hat d_p(\phi_\ast\tilde x_i,\phi_\ast\tilde
	y_i)\lto i0$. We require to show that $d(\tilde x_i,\tilde
	y_i)\lto i0$.  By choosing a subsequence if necessary, we can
	assume that $\phi(x_i)\to \hat x\in \Mhat$, and hence that
	$\phi(y_i)\to \hat x$. From Proposition~\thmref{proper inward and
	outward cones}, if necessary by excluding a finite initial part of
	the sequence and applying a fixed homothety, we can ensure that
	there are inward cones of the form $w_i-\circover C$,
	$z_i-\circover C$, where in the coordinate chart $(U_\alpha,
	p_\alpha)$ thus fixed we write $w_i=p_\alpha(\phi(x_i))$ and
	$z_i=p_\alpha(\phi(y_i))$.  Recall that $K$ is the Lipshitz
	constant in the definition of $\circover C$. Let $\Pi:\R^n\to
	\R^{n-1}$ be the projection onto the first $n-1$ coordinates, and
	let $e$ denote Euclidean distance in the coordinate system now
	chosen. Set
	\begin{eqnarray*}
		e_i\defeq e(w_i,z_i) \\
		\Sigma_i\defeq\R^{n-1}\times\{w_i^n-(K+1)e_i\} \\ 
		P_i\defeq\Pi((w_i-\circover C)\cap\Sigma_i) \\
		Q_i\defeq\Pi((z_i-\circover C)\cap\Sigma_i)
	\end{eqnarray*}
	Then $p_\a(\hat x),z_i,w_i$ are on the same side of the
	hyperplane $\Sigma_i$, $P_i$ is a ball of radius $(1+1/K)e_i$ and
	centre $w_i^\ast\defeq\Pi(w_i)$, while $Q_i$ is a ball of radius
	$((K+1)e_i +z_i^n-w_i^n)/K \geq e_i$ and centre
	$z_i^\ast\defeq\Pi(z_i)$. Since $|w_i^\ast-z_i^\ast| \leq e_i$,
	these balls have a non-empty intersection containng some point
	$f_i*=\Pi(f_i)$, say, where $f_i\defeq(f_i*,w_i^n-(K+1)e_i)$.
	Hence from Lemma~\thmref{hand-wave gap dist approxs Eucl}
	\[
		d(\tilde x_i,\tilde y_i)\leq \mbox{const}\times (e(w_i,
		f_i)+e(f_i,z_i)
		\leq (2K+5)e_i.
	\]
	But from Lemma~\thmref{hand-wave gap dist approxs
	Eucl} again, the far right hand side tends to zero, and so the
	result is proved. 
\end{proof}

Another, perhaps more satisfying, way of stating this result is as an
``imbedding theorem''.  (We use ``imbedding'' rather than ``embedding''
as a mnemonic for ``injection''---there is no topological content to
the stament.)

\begin{corollary}
\thmlabel{imbedding}
	Let $\phi\in\cE[\LL]$ have a Lipschitz boundary.  For any boundary
	point $p\in\d_\phi$, the map
	\[
		\overbar{LM}\to \overbar{L\Mhat_p}:
			\lim_{i\to\infty}(x_i,E_i)
			\mapsto\lim_{i\to\infty}(\phi(x_i),\phi_\ast E_i)
	\]
	is one-one, i.e., it is an imbedding.
\end{corollary}

\section{Application to regular abstract boundaries}

We see that we can neither ``coalesce'' two regular 
boundary points into one without destroying the Lipschitz nature of
the boundary near the points; similarly we cannot ``tear'' one
Lipschitz regular boundary point into two:

\begin{theorem}
\thmlabel{can't split regular boundary points}
	Let $\LL\sub\mathbf C^1$ and $g\in\mathbf C^1$.
	A \rom({\it regular\/}\rom) boundary set $B\sub\d_\phi$,
	$\phi\in\cE[\LL]$, of more than one point cannot be covered by a
	single \rom({\it Lipschitz regular\/}\rom) point
	$q\in\d_\psi$, $\psi\in\cL\cE[\LL]$.
\end{theorem}

\begin{proof}
	Let $p_i$, $i=1,2$, be distinct points of $B$, assumed to be both
	covered by a single boundary point $q\in\d_\psi$.  We first deal
	with the first of the `potential problems' mentioned just before
	\S\secref{after potential problems}.  Of course we can construct a {\em
	single {\rm($C^1$)} coordinate chart\/} $(U,\tau)$ containing the
	two points, using the tubular neighbourhood theorem and a $C^1$
	curve connecting the two points (the latter existing by
	connectedness of $M$).  Since both these points are $C^1$ regular
	boundary points, then we may assume that $U_{p_1}\cap U_{p_2}=\es$
	(c.f.~Definition~\defref{extendibility...}), and thus that there
	is a {\em simultaneous\/} extension of $g$ to
	$\Mhat_{{p_1},{p_2}}\defeq\phi(M)\cup U_{p_1}\cup U_{p_2}$.  We
	may then construct the bundle metric $\hat d_{{p_1},{p_2}}$ as we
	did earlier when considering extensions about a {\em single\/}
	point. 
	(The astute reader will have noted that a version of
		Theorem~\thmref{need lip boundary} using this new bundle
		metric will not be possible in general, for we cannot now
		shrink our neighbourhood to make ``$g_{ij}\approx\eta_{ij}$''.
		Fortunately, we only need to use Theorem~\thmref{need lip
		boundary} about the single (regular) point $q$.
		As a consequence of this \emph{we do not need the boundary
		$\d_\phi$ to be Lipschitz,} but only the boundary $\d_\psi$
		about the single point $q$.)

	Let $\{x_{i,j}\}_{j=1}^\infty$ be sequences of points of $M$ for
	which $\phi(x_{i,j})\stackrel{j\to\infty}\longrightarrow p_i$, and
	$\tilde x_{i,j}$ be points of the fibres over $\phi(x_{i,j})$ in
	$L\Mhat_\phi$.  We shall write $\tilde
	x_{k,j}\defeq(\phi(x_{k,j}),E_{k,j})$, $k=1,2$. Note
	that, as yet, we have said nothing about the fibre elements
	$\mathbf E_{k,j}$, $k=1,2$.

	Now $\psi(x_{i,j}),\,i=1,2$, both
	converge to the covering point $q$. We can choose the fibre
	components of the $\tilde x_{i,j}$ so that these sequences also
	converge to the same point $\tilde q\in L\Mhat_\psi$.  This means
	that $\hat d_q$-distance between the relevant points tends to
	zero, so by Theorem~\thmref{need lip boundary} we must have
	\[
		d\(\tilde x_{1,j},\tilde x_{2,j}\)\lto j0.
	\]
	But $\hat d_{p_1,p_2}(\tilde p_1,\tilde p_2)=2R>0$, so $\hat
	d_{p_1,p_2}(\tilde x_{1,j},\tilde x_{2,j})>R$ for large $j$. Since
	``$\hat d_{p_1,p_2}\le d$'',
	\[
		d\(\tilde x_{1,j},\tilde x_{2,j}\)>R>0
	\]
	for all large enough $j$,   in contradiction.
\end{proof}

We arrive at a result sitting well with intuition.

\begin{corollary}
\thmlabel{two lip regular boundary sets...}
	Let $\LL\sub\mathbf C^1$ and $g\in\mathbf C^1$.
	Lipschitz regularity of boundary sets means that they ``cannot be
	blown up further'' without destroying regularity.  To be more
	precise, consider two regular boundary sets
	$B_1\sub\d_\phi,B_2\sub\d_\psi$ of envelopments
	$\phi,\psi\in\cE[\LL]$. Assume that the first envelopment $\phi$
	has a Lipschitz boundary.  Then no point of $B_1$ can
	cover \textbf{more than} one point of $B_2$.

	If both envelopments have Lipschitz boundaries and $B_1\sim B_2$,
	then each point of $B_1$ is \emph{equivalent to} precisely one
	point of $B_2$.
\end{corollary}
\begin{proof}
	Let $p\in B_1$, and {let $Q\defeq\set q\in B_2 for \hbox{if }
	q=\lim_j\psi(x_j), \hbox{ then } p=\lim_j\phi(x_j)\eset$}. Then
	$Q$ is either empty, in which case $p$ covers no subset of
	$B_2$, or is the maximal subset of $B_2$ covered by $p$.  In
	the latter case we apply Theorem~\thmref{can't split regular
	boundary points}, to get that $Q$ must be a singleton. Thus no
	point of $B_1$ can cover more than one point of $B_2$.

	For the second part, note that by symmetry, the reverse of
	this statement is true, and the result follows.
\end{proof}

The following lemma,  although perhaps rather useless in other
situations, does not {\em per se\/} depend on the Lipschitz, regular
nature of the boundary sets to which we will apply it.

\begin{lemma}
\thmlabel{p->q continuous}
	If $B_1\sim B_2$ and the function $p\mapsto Q$ of
	Corollary~\thmref{two lip regular boundary sets...} is
	singleton-valued \rom(in which case we say $p\mapsto q$, if
	$Q=\{q\}$\rom), and indeed maps each $p$ to an equivalent $q$,
	then $B_1\to B_2:p\mapsto q$ is continuous.
\end{lemma}
\begin{proof}
	If not, then there is an open neighbourhood $V$ of $q\in B_2$ such
	that there is no neighbourhood $U$ of $p$ all of whose boundary
	points correspond to points of $V$. If there are small enough
	neighbourhoods $U$ intersecting $B_1$ only at $p$, then we get a
	trivial contradiction.  So there is a sequence of boundary points
	$p_i\in B_1$, equivalent to points {\em outside\/} $V$, which
	converge to $p$.

	But $q$ covers $p$, so there is a sub-neighbourhood $U'$ of $U$
	such that $\psi\circ\inv\phi(U'\cap\phi(M))\sub V$. {\em All\/}
	sequences of points $\{x^i_j\}_j\sub\inv\phi(U'\cap\phi(M))$, with
	respective images under $\phi$ converging in $j$ to the respective
	$p_i\in B_1$, must have images under $\psi$ which converge to
	points outside of $V$.  However, the ``diagonal'' sequence
	$\{x_i^i\}_i$ has $\phi(x_i^i)\lto ip$, and so $\psi(x_i^i)\lto
	iq$, in contradiction since $\Mhat_\psi\sm V$ is closed.
\end{proof}

By symmetry, we have the following.

\begin{theorem}{\bf (Characterization of Lipschitz regular abstract boundary,
smooth case)\ }
\thmlabel{equivalent Lipschitz regular boundary sets homeomorphic}
	If $M$ is a differentiable manifold and $g\in
	\LL\sub\mathbf C^1$, then all representatives of an
	equivalence class of Lipschitz regular boundary sets
	$[B]\in\cL\cR*[M,g,\L]$ are homeomorphic.  That is, ``the''
	topology of an equivalence class of boundary sets is independent
	of the particular envelopment used to define it, if all
	envelopments considered have Lipschitz, regular boundaries.
\end{theorem}

Note that this means that every representative of a Lipschitz
$C^1$-regular abstract boundary point $[p]\in\cL\cR[M,g,\L]$ is itself
a point.  Of course, the nature of a Lipschitz boundary forces the
original point $p$ to be a point of an ($n-1$)-dimensional boundary
hypersurface.  As was alluded to earlier, it says that any envelopment
of the space-time in which our original point is ``blown up'' to be
represented by anything other than a single point destroys $ C^1$
regularity of at least part of the new representative, {\em whether
this new boundary is Lipschitz or not}.

This gives us a rather complete characterisation of those sections of
boundaries of (e.g.) space-times through which the metric can be
extended in a smooth ($C^1$) fashion.  We are on firm {\em
mathematical}, if not physical, ground when we suggest that each point
of a Lipschitz of boundary of a space-time through which the metric
can be extended smoothly, has something to distinguish it from all
others.

It is a natural question to ask whether the same (or a similar)
characterisation is valid when we relax the smoothness assumption
somewhat, allowing, e.g., gravitational shock waves.  After all, we
would certainly hope that the presence of gravitationl waves in the
space-time and extension thereof cannot ``distort'' the boundary so
that the topological type of a portion of ``boundary hypersurface'',
through which extension is possible, could depend on the precise way
in which the extension is performed.

Since boundaries are of physical interest precisely in situations of
collapse, where strong gravitational fields are liable to generate
shock- and impulse-waves, it is important to be able to relax as far
as possible the smoothness assumptions in this work. It would be a
problem indeed for our regular abstract boundaries if it turned out
that ``non-smooth perturbations'' of a smooth metric could result in a
space-time where, unlike the smooth case, the topological type of
a portion of ``regular boundary hypersurface'' could depend on
the precise way in which the extension is performed. In a forthcoming
paper we will extend our results in this direction.

\section*{Acknowledgements}

CJF would like to thank the School of Mathematical Sciences at the
Australian National University, particularly Profs.~Derek~W.~Robinson
and Neil~S.~Trudinger, for sending him to Southampton, where most of
this collaboration was undertaken.  He also thanks
Drs.~Marcus~Kriele and Giselle~C.~Lim for their generous and
invaluable input.

\section*{References}
%\bibliographystyle{plain}
%\bibliography{pabbrev,mathphys,a-bnd,gr,algebra,diff}

\begin{thebibliography}{10}

\bibitem{Scott94}
Scott S M and Szekeres P
1994% The abstract boundary---a new approach to singularities of manifolds.
 {\em J. Geom. Phys.} {\bf 13}~223--253.

\bibitem{Fama95}
Fama C J  and Scott S M.
%\newblock Invariance properties of boundary points of open embeddings of
%  manifolds and their application to the abstract boundary.
1995
\emph{Proc.~AMS-CMS special session on Geometric Methods in Mathematical
Physics}
(American  Mathematical Society)
pp~79--111.

\bibitem{O'Neill83}
O'Neill B
1983
\newblock {\em Semi-Riemannian geometry: with applications to relativity},
(Pure and applied mathematics vol 103)
\newblock (New York: Academic Press).

\bibitem{Schmidt71}
Schmidt B G.
1971 %A new definition of singular points in general relativity.
 {\em Gen. Rel. Grav.} {\bf 1}~269--280.

\bibitem{Clarke93b}
Clarke C J S
1993
\newblock {\em The analysis of space-time singularities} 
  Cambridge lecture notes in physics vol~1
\newblock (Cambridge University Press).

\bibitem{Dodson78}
Dodson C T J
1978 %Space-time edge geometry.
 {\em Int.. J. Theoret. Phys.} {\bf 17}~389--504.

\bibitem{Dodson80}
Dodson C T J
1980
\newblock {\em Categories, Bundles and Spacetime Topology}, 
  Shiva Mathematics Series vol~1
\newblock (Shiva Publishing Ltd.).

\bibitem{Geroch72}
Geroch R, Kronheimer E H, and Penrose R
1972 Ideal points in space-time.
  \PRS{\em London Ser. A} {\bf 327}~545--567.

\bibitem{Beem81}
Beem J K and Ehrlich P E
1981
\newblock {\em Global {L}orentzian geometry}, Pure and
  applied mathematics vol~67
\newblock (New York: M. Dekker).

\bibitem{Hughes76}
Hughes T~J~R, Kato T , and Marsden J E
1976
%\newblock Well-posed quasi-linear second-order hyperbolic systems with
%  applications to nonlinear elastodynamics and general relativity.
\newblock {\em Arch. Rational Mech. Anal.}, {\bf 63(3)} 273--294.

\bibitem{Geroch87}
Geroch R  and Traschen J
 1987
%\newblock Strings and other distributional sources in general relativity.
\PR{\em D}.

\bibitem{Hirsch76}
Hirsch M W
1976
\newblock {\em Differential topology}, Graduate texts in
  mathematics  vol~33
\newblock (Springer-Verl{\"a}g).

\bibitem{Hartman64}
Hartman P
 1964
\newblock {\em Ordinary Differential Equations}
\newblock (John Wiley and Sons, Inc.).
\end{thebibliography}

\end{document}